\documentstyle[12pt,epsfig,axodraw,a4]{article} 
\def\bild#1#2{    
        \vspace*{-5mm}
        \begin{center}
        \begin{math}
        \epsfxsize#2cm
        \epsffile{#1}
        \end{math}
        \end{center}  }
\newcommand{\vs}{\vspace{-0.25cm}}
\begin{document} 
\begin{center}
\large{\bf Chiral 2\begin{boldmath}$\pi$\end{boldmath}-exchange NN-potentials:
Relativistic \begin{boldmath}$1/M^2$\end{boldmath}-Corrections}

\bigskip

N. Kaiser\\

\medskip

{\small Physik Department T39, Technische Universit\"{a}t M\"{u}nchen,
    D-85747 Garching, Germany}

\end{center}

\medskip

\begin{abstract}
We calculate in baryon chiral perturbation theory the relativistic
$1/M^2$-corrections to the leading order two-pion exchange diagrams. We give 
explicit expressions for the corresponding one-loop NN-amplitudes in momentum
space. The resulting isovector central and isoscalar spin-spin and tensor 
NN-amplitudes involve non-static terms proportional to the squared nucleon 
center-of-mass momentum $p^2$. From the two-pion exchange box diagrams we 
obtain an isoscalar quadratic spin-orbit NN-amplitude. We give also analytical 
expressions for the corresponding NN-potentials in coordinate space. The 
diagrammatic results presented here make the chiral NN-potential complete 
at next-to-next-to-next-to-leading order.

\end{abstract}


\bigskip
\bigskip

Over the past years effective field theory methods have been successfully
applied to the two-nucleon system at low and intermediate energies 
\cite{weinb,kolck,kaplan,epelb,nnpap1,nnpap2}. The idea of constructing
the NN-potential from effective field theory was put forward by Weinberg
\cite{weinb} and this was first taken up by van Kolck and collaborators 
\cite{kolck} who used "old-fashioned" time-ordered perturbation theory. Later,
the systematic method of unitary transformations was employed by Epelbaum, 
Gl\"ockle and Mei{\ss}ner \cite{epelb} to construct an energy-independent 
NN-potential from the effective chiral pion-nucleon Lagrangian at 
(next-to-)next-to-leading order. Based on one- and two-pion exchange and nine 
adjustable NN-contact interactions which contribute only to S- and P-waves a
good description of the deuteron properties as well as the NN phase-shifts and
mixing angles below $T_{\rm lab}=300$\,MeV was found in that framework 
\cite{epelb}. Also recently, the elastic proton-proton scattering database 
below $350\,$MeV laboratory kinetic energy (consisting of 1951 data points) 
has been analyzed in terms of $1\pi$-exchange and chiral $2\pi$-exchange at 
next-to-next-to-leading order in ref.\cite{nijmeg}. The resulting good
$\chi^2/{\rm dof}\leq 1$ constitutes a convincing proof for the presence of the
chiral $2\pi$-exchange in the long-range proton-proton strong interaction. It
was concluded in ref.\cite{nijmeg} that $1\pi$-exchange together with chiral
$2\pi$-exchange gives a very good NN-force at least as far inwards as
$r=1.4\,$fm internucleon distance. All shorter range components of the 
NN-interaction have been effectively parametrized in ref.\cite{nijmeg} by 23 
boundary condition parameters.

At present, there is much interest in extending the calculations of the 
two-nucleon  system \cite{mach} (and also the analysis of the NN database) to
one order higher in the chiral expansion. This requires the full knowledge of
the next-to-next-to-next-to-leading order (N$^3$LO) chiral NN-potential. In 
momentum space this corresponds to terms in the NN T-matrix which are of fourth
order in small external momenta and the pion mass, denoted generically by 
${\cal O}(Q^4)$. In particular, all two-loop diagrams of the process $NN\to NN$
with leading order vertices contribute at that order, ${\cal O}(Q^4)$. The
chiral $3\pi$-exchange has recently been calculated completely in  
ref.\cite{3pi}. Furthermore, results for the two-loop $2\pi$-exchange are now 
available from ref.\cite{twoloop}. In that work also the one-loop contributions
(of order ${\cal O}(Q^4)$) proportional to the second and third order 
low-energy constants $c_{1,2,3,4}$ and $\bar d_j$, related to elastic $\pi 
N$-scattering, have been evaluated. The only remaining and presently unknown
pieces of the NN T-matrix at order ${\cal O}(Q^4)$ are the relativistic 
$1/M^2$-corrections to the leading order one-loop $2\pi$-exchange. The
purpose of this work is to present analytical results for these remaining
contributions and thus to complete the chiral NN-potential at
next-to-next-to-next-to-leading order.  

Let us first give some basic definitions in order to fix our notation. We 
are considering elastic NN-scattering in the center-of-mass frame, i.e. the 
process $N_1(\vec p\,) +N_2(-\vec p\,) \to  N_1(\vec p+\vec q \,)+N_2(-\vec p-
\vec q\,)$. The corresponding on-shell T-matrix has the following general form,
\begin{eqnarray} {\cal T}_{NN} &=& V_C+ \vec \tau_1 \cdot \vec \tau_2  W_C +
\big[V_{S}+ \vec \tau_1 \cdot \vec \tau_2  W_{S} \big]\,\vec\sigma_1
\cdot \vec \sigma_2+ \big[ V_T + \vec \tau_1 \cdot \vec \tau_2 W_T 
\big]\,  \vec  \sigma_1 \cdot \vec q \, \vec \sigma_2 \cdot \vec q \nonumber 
\\ & &+ \big[ V_{SO}+\vec\tau_1 \cdot \vec \tau_2 W_{SO} \big] \,i( 
\vec \sigma_1 +\vec  \sigma_2)\cdot (\vec q\times \vec p\,) \nonumber 
\\ & & + \big[ V_Q+ \vec \tau_1 \cdot \vec \tau_2 W_Q \big]\,\vec\sigma_1
\cdot(\vec q\times  \vec p\,) \,\vec \sigma_2 \cdot(\vec q\times \vec p\,)\,,
\end{eqnarray}
with $q=|\vec q\,|$ the momentum transfer between the initial and final-state 
nucleon, subject to the constraint $2\vec p\cdot \vec q = -q^2$. The subscripts
$C,S,T,SO$ and $Q$ refer to the central, spin-spin, tensor, spin-orbit and
quadratic spin-orbit components, each of which occurs in an isoscalar 
($V_{C,\dots,Q}$) and an isovector ($W_{C,\dots,Q})$ version. Furthermore, 
$\vec \sigma_{1,2}$ and $\vec \tau_{1,2}$ in eq.(1) denote the spin and
isospin operators of the two nucleons. 

\bigskip

\bigskip

\bild{rel2fig1.epsi}{2.5}
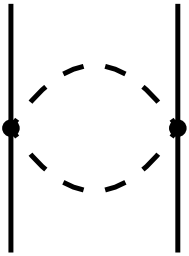
{\it Fig.1: $2\pi$-exchange bubble diagram. The combinatoric factor of this 
graph is 1/2.}

\bigskip

\bigskip

Let us now turn to the results for the $1/M^2$-corrections to the chiral 
$2\pi$-exchange. The bubble or football diagram involving two isovectorial 
Tomozawa-Weinberg vertices is shown in Fig.\,1. A convenient way to obtain 
all $1/M^2$-corrections from such a one-loop diagram goes as follows. At the 
left nucleon line one expands in the NN center-of-mass frame the pertinent 
relativistic $\pi\pi NN$-vertex sandwiched between Dirac-spinors for the out- 
and in-going nucleon in powers of $1/M$ up to quadratic order, i.e. ${\cal O}
(M^{-2})$. The analogous expression at the right nucleon line is obtained
simply by switching the signs of three-momenta $\vec p$ and $\vec q$ and of the
loop four-momentum $(l_0, \vec l\,)$. Of course, one also relabels the spin and
isospin operators, $\vec \sigma_1 \to \vec \sigma_2$ and $\vec \tau_1 \to \vec
\tau_2$. In a last step, one multiplies the product of these both expressions
with the pion propagators and carries out the four-dimensional loop
integration. The total result for the $1/M^2$-corrections to the bubble diagram
in Fig.\,1 reads  (displaying only non-vanishing contributions),    
\begin{equation} W_C = {(q^2-4p^2) w^2 \, L(q) \over 768\pi^2 M^2 f_\pi^4}\,,
 \end{equation} 
\begin{equation} W_{SO} =3W_T = -{3\over q^2} W_S= - {w^2 \, L(q) \over 512
\pi^2 M^2 f_\pi^4}\,, \end{equation}
with the frequently occuring logarithmic loop function
\begin{equation} L(q) = {w\over q} \ln{w+q \over 2m_\pi} \,, \qquad
w=\sqrt{4m_\pi^2+q^2}\,. \end{equation}
In eqs.(2,3) and in the following we omit purely polynomial terms which can be
absorbed in the strength of some zero-range NN-contact interactions. In fact, 
we are interested here only in non-polynomial or finite-range terms from chiral
$2\pi$-exchange. Note also that the isovector central NN-amplitude $W_C$ in 
eq.(2) has a non-static contribution proportional the squared nucleon 
center-of-mass momentum $p^2$.

\bigskip

\bigskip

\bild{rel2fig2.epsi}{7}
{\it Fig.2: $2\pi$-exchange triangle diagrams.}

\bigskip

\bigskip

Fig.\,2 shows the $2\pi$-exchange triangle diagrams. The expression for the 
$\pi\pi NN$-vertex at the left (or right) nucleon line of the first (second)
diagram can be taken over from the calculation of the bubble diagram. At the 
other (right or left) nucleon line one expands in powers of $1/M$ the 
product of upper pseudovector $\pi N$-vertex, nucleon Dirac-propagator and 
lower pseudovector $\pi N$-vertex sandwiched between out- and in-going 
Dirac-spinors. Collecting all possible terms proportional to $1/M^2$ and 
performing the loop integration over the pion propagators, the total result for
the $1/M^2$-corrections to the triangle diagrams in Fig.\,2 reads     
\begin{equation} W_C = {g_A^2 \, L(q) \over 96\pi^2 M^2 f_\pi^4}\bigg\{{11\over
4} q^4 +5m_\pi^2 q^2 +3m_\pi^4-6m_\pi^6 w^{-2} -p^2(8m_\pi^2 +5q^2)
\bigg\}\,,\end{equation}  
\begin{equation} W_T = -{1\over q^2} W_S={g_A^2 \, L(q) \over 48\pi^2 M^2 
f_\pi^4} \bigg\{m_\pi^2+{7\over 16}q^2 \bigg\} \,,\end{equation}
\begin{equation} W_{SO} ={g_A^2 \, L(q) \over 32\pi^2 M^2 f_\pi^4} \bigg\{
m_\pi^2+{3\over 8}q^2\bigg\} \,.\end{equation}
Again, the isovector central NN-amplitude $W_C$ in eq.(5) has a non-static
piece proportional to $p^2$. In the heavy nucleon mass limit $M\to \infty$ the 
diagrams in Fig.\,1 and Fig.\,2 give rise only to a non-vanishing isovector
central NN-amplitude $W_C$ (see eq.(14) in ref.\cite{nnpap1}) and one finds 
that the non-static $p^2$-terms in eqs.(2,5) are in fact proportional to it.

\bigskip

\bigskip

\bild{rel2fig3.epsi}{7}
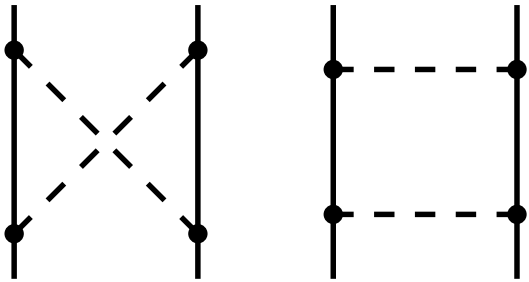
{\it Fig.3: $2\pi$-exchange box diagrams. The (right) planar box graph includes
the iterated $1\pi$-exchange.}

\bigskip

\bigskip

Fig.\,3 shows the $2\pi$-exchange crossed box and planar box diagrams. The
crossed box diagram (of isospin structure $3+2 \vec \tau_1 \cdot \vec \tau_2$)
can be evaluated straightforwardly by making use of the expression for the 
sequential pion absorption and emission process derived in the calculation of 
the triangle diagrams. At the left and the right nucleon line these expressions
differ just by the sign of the three-momenta $\vec p$ and $\vec q$. The planar
box diagram (of isospin structure $3-2 \vec \tau_1 \cdot \vec\tau_2$) requires
some special treatment since it contains the so-called iterated $1\pi
$-exchange. As outlined in section 4.2 of ref.\cite{nnpap1} one must perform 
the $dl_0$ loop integration (using e.g. residue calculus) before the 
$1/M$-expansion in order to isolate the iterated $1\pi$-exchange together with 
its own relativistic $1/M$-corrections. From that (and also from a
consideration of the Lorentz-invariant NN two-body phase space) one finds a 
dependence of the iterated $1\pi$-exchange on the nucleon mass $M$ of the form 
$M^2/\sqrt{M^2+p^2}= M-p^2/2M+{\cal O}(M^{-3})$. Since there obviously is no 
contribution to the iterated $1\pi$-exchange proportional to $1/M^2$ one can
treat (for the present purpose) the planar box graph in the same way as the
crossed box graph. Namely, one performs first the $1/M$-expansion of vertices
and propagators at the left and the right nucleon line and then one carries out
the four-dimensional loop integration. We have used this method as well as the 
continuation of the method outlined in section 4.2 of ref.\cite{nnpap1} (where
things are basically done in the reverse order) in order to calculate the 
$1/M^2$-corrections to the planar box graph and we have found perfect
agreement. Sorting finally the contributions from the crossed 
box diagram and planar box diagram in Fig.\,3 into isoscalar and isovector 
components, we find after a tedious calculation 
\begin{equation} V_C = {g_A^4 \over 32\pi^2 M^2 f_\pi^4}\bigg\{ {1\over 2}
m_\pi^6 w^{-2} +\Big[2m_\pi^8 w^{-4}+8m_\pi^6 w^{-2} -q^4 -2m_\pi^4\Big] L(q)
\bigg\}\,,\end{equation}
\begin{eqnarray} W_C &=& {g_A^4 \over 192\pi^2 M^2f_\pi^4}\bigg\{4m_\pi^6w^{-2}
+\bigg[p^2\Big(20m_\pi^2+7q^2-16m_\pi^4 w^{-2}\Big) \nonumber \\ && +16m_\pi^8
w^{-4}+12 m_\pi^6 w^{-2} -{27\over 4}q^4 -11m_\pi^2 q^2-6m_\pi^4 \bigg] L(q) 
\bigg\}\,,  \end{eqnarray}
\begin{equation} V_T = -{1\over q^2} V_S=- {g_A^4 \, L(q) \over 32\pi^2 M^2 
f_\pi^4} \bigg\{p^2+{3\over 8} q^2 +m_\pi^4 w^{-2} \bigg\} \,,\end{equation}
\begin{equation} W_T = -{1\over q^2} W_S=- {g_A^4 \, L(q) \over 384\pi^2 M^2 
f_\pi^4} \bigg\{7m_\pi^2+{17\over 4} q^2 +4m_\pi^4 w^{-2} \bigg\} \,,
\end{equation}
\begin{equation} V_{SO} ={g_A^4 \, L(q) \over 8\pi^2 M^2 f_\pi^4} \bigg\{ {11
\over 32} q^2 +m_\pi^4 w^{-2}\bigg\} \,,\end{equation}
\begin{equation} W_{SO} ={g_A^4 \, L(q) \over 384\pi^2 M^2 f_\pi^4} \bigg\{
4m_\pi^4 w^{-2}-{11\over 4}q^2 -9m_\pi^2 \bigg\} \,,\end{equation}
\begin{equation} V_Q =-{g_A^4 \, L(q) \over 32\pi^2 M^2 f_\pi^4}\,.
\end{equation} 
Here, one observes that the isovector central NN-amplitude $W_C$ and the
isoscalar spin-spin and tensor NN-amplitudes $V_{S,T}$ in eqs.(9,10) receive 
non-static contributions proportional to $p^2$. Another interesting feature is
the occurrence of an isoscalar quadratic spin-orbit NN-amplitude $V_Q$ in 
eq.(14), whereas the analogous isovector NN-amplitude vanishes, $W_Q=0$. This 
completes the presentation of analytical results for the momentum space 
NN-amplitudes.    

For certain applications \cite{nijmeg} it is useful to have a representation 
of the NN-interaction in terms of local coordinate space potentials. These
potentials are related to the momentum space NN-amplitudes by a negative
Fourier-transform, $-(2\pi)^{-3} \int d^3 q\,\exp(i\vec q \cdot \vec r\,)$. 
The isoscalar central, spin-spin, tensor and spin-orbit potentials, denoted
here by $\widetilde V_{C,S,T,SO} (r)$, are accompanied by the operators $1$,
$\vec \sigma_1\cdot\vec \sigma_2$, $3\,\vec\sigma_1\cdot \hat r\,\vec \sigma_2 
\cdot \hat r - \vec \sigma_1\cdot \vec \sigma_2$ and $-{i\over 2} (\vec 
\sigma_1+\vec \sigma_2) \cdot (\vec r \times \vec \nabla)$. In the case of the 
isovector potentials $\widetilde W_{C,S,T,SO}(r)$ an additional isospin 
operator $\vec \tau_1 \cdot \vec \tau_2$ occurs. The finite-range parts
(disregarding zero-range $\delta^3(\vec r\,)$-terms) of the coordinate space
potentials following from eqs.(2-14) are most conveniently obtained through
their spectral-representations \cite{3pi,twoloop} with the help of the formula 
Im\,$L(i\mu)= -({\pi/2\mu})\sqrt{\mu^2-4m_\pi^2}$. It turns out that all
these potentials can be written analytically in terms of two modified 
Bessel-functions  $K_{0,1}(2x)$, with the dimensionless variable $x=m_\pi r$. 
We list them separately for the three classes of diagrams. 

\medskip

\noindent
i) Bubble diagram in Fig.\,1: 
\begin{equation} \widetilde W_C(r) = {m_\pi \over 128\pi^3 M^2 f_\pi^4 r^6}
\Big\{ 2x(5+x^2+p^2r^2)K_0(2x)+(10+7x^2+2p^2r^2)K_1(2x)\Big\}\,,\end{equation} 
\begin{equation} \widetilde W_S(r) = {m_\pi \over 384\pi^3 M^2 f_\pi^4 r^6}
\Big\{ 2x(5+x^2)K_0(2x)+(10+7x^2)K_1(2x)\Big\}\,,\end{equation}
\begin{equation} \widetilde W_T(r) = -{m_\pi \over 1536\pi^3 M^2 f_\pi^4 r^6}
\Big\{ x(35+4x^2)K_0(2x)+(35+20x^2)K_1(2x)\Big\}\,,\end{equation}
\begin{equation} \widetilde W_{SO}(r) =-{3m_\pi \over 256\pi^3 M^2 f_\pi^4 r^6}
\Big\{ 5x K_0(2x)+(5+2x^2)K_1(2x)\Big\}\,.\end{equation}

\noindent
ii) Triangle diagrams in Fig.\,2:
\begin{eqnarray} \widetilde W_C(r) &=&{g_A^2 m_\pi\over 64\pi^3 M^2 f_\pi^4r^6}
\Big\{ x(110+34x^2+x^4+p^2r^2)K_0(2x)\nonumber \\ &&+[110+89x^2+9x^4+(10+4x^2)
p^2r^2]K_1(2x)\Big\}\,,\end{eqnarray}
\begin{equation} \widetilde W_S(r) =-{g_A^2m_\pi \over 192\pi^3 M^2f_\pi^4 r^6}
\Big\{ 10x(7+2x^2)K_0(2x)+(70+55x^2+4x^4)K_1(2x)\Big\}\,,\end{equation}
\begin{equation} \widetilde W_T(r) ={g_A^2m_\pi\over 768\pi^3 M^2 f_\pi^4 r^6}
\Big\{ x(245+52x^2)K_0(2x)+(245+170x^2+8x^4)K_1(2x)\Big\}\,,\end{equation}
\begin{equation} \widetilde W_{SO}(r)={g_A^2m_\pi\over 128\pi^3M^2 f_\pi^4 r^6}
\Big\{x(45+4x^2) K_0(2x)+(45+24x^2)K_1(2x)\Big\}\,.\end{equation}

\noindent
iii) Box diagrams in Fig\,.3:
\begin{equation} \widetilde V_C(r) =-{g_A^4m_\pi\over 16\pi^3 M^2 f_\pi^4 r^6}
\Big\{x(30+12x^2+x^4)K_0(2x)+\Big(30+27x^2+{9\over 2}x^4+{x^6\over 8}\Big)
K_1(2x)\Big\}\,,\end{equation}
\begin{eqnarray} \widetilde W_C(r) &=&{g_A^4m_\pi\over 384\pi^3M^2 f_\pi^4 r^6}
\Big\{ x[p^2r^2(8x^2-42)-810-258x^2-6x^4]K_0(2x) \nonumber \\ && -[810+663x^2
+70x^4 +4x^6+p^2r^2(42+8x^2)] K_1(2x)\Big\}\,,\end{eqnarray}
\begin{eqnarray} \widetilde V_S(r) &=&{g_A^4m_\pi \over 192\pi^3 M^2f_\pi^4r^6}
\Big\{x(90+36x^2+4x^4-12p^2r^2)K_0(2x)\nonumber \\ && +[90+81x^2+14x^4-p^2r^2
(12+8x^2)]K_1(2x)\Big\}\,,\end{eqnarray}
\begin{equation} \widetilde W_S(r) ={g_A^4m_\pi\over 1152\pi^3 M^2 f_\pi^4 r^6}
\Big\{x(510+162x^2+8x^4)K_0(2x)+(510+417x^2+44x^4)K_1(2x)\Big\}\,,
\end{equation}
\begin{eqnarray} \widetilde V_T(r) &=&{g_A^4m_\pi \over768\pi^3 M^2f_\pi^4 r^6}
\Big\{x(48p^2r^2-315-114x^2-8x^4)K_0(2x)\nonumber \\ && +[p^2r^2(60+16x^2)-315
-270x^2-40x^4] K_1(2x) \Big\}\,,\end{eqnarray}
\begin{equation} \widetilde W_T(r) =-{g_A^4m_\pi\over 4608\pi^3 M^2f_\pi^4 r^6}
\Big\{x(1785+456x^2+16x^4)K_0(2x)+(1785+1320x^2+112x^4)K_1(2x)\Big\}\,,
\end{equation}
\begin{equation} \widetilde V_{SO}(r) ={g_A^4m_\pi\over 128\pi^3M^2f_\pi^4 r^6}
\Big\{x(165+52x^2)K_0(2x)+(165+132x^2+16x^4)K_1(2x) \Big\}\,,\end{equation}
\begin{equation} \widetilde W_{SO}(r) ={g_A^4m_\pi\over 768\pi^3M^2f_\pi^4 r^6}
\Big\{-x(165+4x^2)K_0(2x)+(8x^4-78x^2-165)K_1(2x) \Big\}\,,\end{equation}
\begin{equation} \widetilde V_Q(r) ={g_A^4m_\pi\over 64\pi^3M^2f_\pi^4 r^6}
\Big\{12x K_0(2x)+(15+4x^2)K_1(2x) \Big\}\,,\end{equation}
The last potential $ \widetilde V_Q(r)$ is accompanied by the quadratic
spin-orbit operator $-\vec \sigma_1\cdot (\vec r \times \vec \nabla)\, \vec
\sigma_2 \cdot (\vec r \times \vec \nabla)$. In order to Fourier-transform the
$w^{-4} L(q)$-terms in eqs.(8,9) it is useful to exploit the relation $4m_\pi^2
w^{-4} L(q) + w^{-2} =-m_\pi ({\partial/\partial m_\pi})[w^{-2}L(q)]$.

In Table\,1, we present numerical values for the coordinate space potentials
eqs.(15-31) at $p=0$, using the parameters $g_A=1.3$, $m_\pi=138\,$MeV,
$f_\pi=92.4\,$MeV and $M=939\,$MeV. One observes that the attractive central 
potentials, $\widetilde V_C(r)$ and $\widetilde W_C(r)$, and the repulsive 
isoscalar spin-orbit potential $\widetilde V_{SO}(r)$ are quite sizeable. The
inclusion of the relativistic $1/M^2$-corrections to the chiral $2\pi$-exchange
in NN phase-shift calculations etc. may therefore not be just a small effect. 
Of course, a firm conclusion about their relevance can only be drawn from the 
complete N$^3$LO chiral NN-potential.  
 
In summary, we have completed here the chiral NN-potential at
next-to-next-to-next-to-leading order by evaluating the relativistic
$1/M^2$-corrections to $2\pi$-exchange. The analytical results presented here 
are in a form such that they can be easily implemented in a N$^3$LO calculation
of the two-nucleon system or in an empirical analysis of low-energy elastic
NN-scattering.  Work along this line is in progress \cite{mach}. With such 
calculations one will be able to determine the maximal range of validity of
chiral multi-pion exchange in the NN-interaction.  

\begin{table}[hbt]
\begin{center}
\begin{tabular}{|c|cccccccc|}
    \hline

    $r$~[fm]&0.7 &0.8&0.9&1.0&1.1&1.2&1.3 &1.4\\ \hline
$\widetilde V_C$~[MeV]
&-185.5& -72.3 &-31.4 &-14.9&-7.56 & -4.06 & -2.29& -1.34\\ 
 $\widetilde W_C$~[MeV]
&-103.7& -40.2 &-17.4 &-8.18&-4.13 &-2.20& -0.28&-0.72 \\ 
 $\widetilde V_S$~[MeV]
&46.4& 18.1 &7.86 &3.73 &1.89& 1.02&0.57& 0.34 \\     
 $\widetilde W_S$~[MeV]
& 23.1 &8.95 &3.87 &1.83&0.92& 0.49&0.28&0.16  \\ 
$\widetilde V_T$~[MeV]
&-40.2& -15.6 &-6.77&-3.20&-1.62&-0.87& -0.49&-0.29 \\ 
 $\widetilde W_T$~[MeV]
& -19.9& -7.69 &-3.32&-1.56&-0.78&-0.42&-0.23&-0.13  \\  
 $\widetilde V_{SO}$~[MeV]
&124.8& 48.3 &20.9 &9.81&4.94& 2.63&1.47&0.85 \\ 
$\widetilde W_{SO}$~[MeV]
& -2.15& -0.78 &-0.31&-0.13&-0.06&-0.028& -0.013&-0.006 \\ 
 $\widetilde V_Q$~[MeV]
& 19.4&7.26 &3.02 &1.37&0.66&0.34&0.18&0.10 \\  
  \hline
  \end{tabular}
\end{center}
{\it Tab.1: Numerical values of the relativistic $1/M^2$-corrections to the 
chiral $2\pi$-exchange NN-potentials (at $p=0$) versus the nucleon distance 
$r$. The units of  these potentials are MeV.}
\vspace{-0.2cm}

\end{table}

\end{document}